\documentclass[aps,prb,twocolumn,groupedaddress,floatfix,showpacs,endfloats]{revtex4-1}
\usepackage{graphicx}
\usepackage{amsmath}
\usepackage{color}
\begin{document}

%\preprint{}

\title{Excitons in Cu$_2$O - from Quantum Dots to Bulk Crystals and\\ Additional Boundary Conditions for Rydberg Exciton - Polaritons}


\author{David Ziemkiewicz}
\email{david.ziemkiewicz@utp.edu.pl}

\author{Karol Karpi\'{n}ski}
\author{Gerard Czajkowski}\author{Sylwia
Zieli\'{n}ska-Raczy\'{n}ska}
%\thanks{}
 \affiliation{Institute of
Mathematics and Physics, UTP University of Science and Technology,
\\Aleje Prof. S. Kaliskiego 7, 85-789 Bydgoszcz, Poland.}


\date{\today}
\begin{abstract}
We propose  schemes for calculation of optical functions of a
semiconductor with Rydberg excitons for a wide interval of
dimensions. We have started with a zero-dimensional structure (Quantum
Dot), then going to one-dimensional (Quantum Wire),
two-dimensional (Quantum Wells and Wide Quantum Wells)
ending on 3-dimensional bulk crystals; our analytical fidings are illustrated numerically showing an agreement with avaliable experimental data. The calculations including
excitons-polaritons  are performed;   the case of large number of
polariton branches is discussed and obtained theoretical
absorption spectra show good agreement with experimental data.
\end{abstract}
\pacs{71.35.-y,78.20.-e,78.40.-q}
\maketitle

\section{Introduction}
Since 2014 the Rydberg excitons (RE) in cuprous oxide, first observed by Kazimierczuk \emph{et al} \cite{Kazimierczuk}, have been subject of extensive
research. Unusual properties of RE \cite{Hecktoeter_2017,Assmann_symmetry} manifested in their interaction with external fields below \cite{Thewes,Schoene,FS95,Magnetoexcitons_2019} and above the gap energy \cite{FK} in linear and nonlinear \cite{Walther,Nonlinear} regime have been studied, both
experimentally \cite{Kazimierczuk,FS96,Heckotter_plasma} and theoretically \cite{Zielinska.PRB,Zielinska.PRB.2016.b,Zielinska.PRB.2016.c}, the list being far from completeness.

In recent years, there has been a dedicated effort  to describe the spectroscopic and optical properties of RE and several methods have been applied. Calculations based on the group theory have been used to obtain the dependence of the spectra on the geometry of external fields \cite{FS95, FS96}
 for RE up to $n=5$, while application of the mesoscopic Real Density Matrix Approach (RDMA) has turned out to be fruitful for description of optical function of semiconductor crystals including RE  for the case of indirect interband transitions, as it was shown in the series of papers by S. Zieli\'{n}ska-Raczy\'{n}ska \emph{et al}. \cite{Zielinska.PRB,Zielinska.PRB.2016.b,Zielinska.PRB.2016.c}
This approach has turned out to be very flexible and general; it allows one to obtain detailed description of RE resonances in various external field configurations and for all excitonic states.
 This, in turn, provides data necessary for potential implementations of RE such as high power excitonic masers
 \cite{maser2} and tunable electro-modulator. \cite{Modulator}

The majority of papers on RE in Cu$_2$O considered RE in bulk crystals or in plane-parallel slabs with dimensions much greater than the incident wave length, and the effective Bohr radius. However,
cuprous oxide nanostructures have recently received attention. \cite{Naka} It seems that quantum-confined structures with RE may be of interest both to research scientists who would be able to explore uncharted areas of fundamental physics of RE in semiconductors in confined geometry and to engineers who might use their unique properties for device applications in the future, paving the way for a whole new class of apparatus such as detectors and optoelectronic switches. The growing interest on optical properties of low dimensional systems (LDS), such as quantum wells, wires and dots, with Rydberg excitons  is noticeable \cite{Takahata, Konzelmann}.  Takahata et al. \cite{Takahata} have begun the studies on RE in low dimension structures  performing the observations of nonlocal response of weakly confined RE in plane-parallel Cu$_2$O films, the  thickness of which ranging from 16 nm to 2000 nm, which are much smaller than those from first experiments i.e.,  in
Ref.\cite{Kazimierczuk,FS96}, where the bulk dimension was around 30-50 $\mu$m. Konzelmann et al. \cite{Konzelmann} has studied theoretically  the optical properties of LDS with Rydberg excitons, focusing their attention on the impact of confinement potentials on the energy shifts of RE in Cu$_2$O LDS. Inspired by these novel LDS in Cu$_2$O, we aim to analyse their optical properties, taking into account multiple Rydberg states.

Quantum size effects become important when the thickness of the
layer $L$ becomes comparable with the de Broglie wavelength of the
electrons or holes. The structures with quantum-confinement
effects  include zero- dimensional Quantum Dots
(QD), one- dimensional Quantum Wires (QWW), and
two-dimensional Quantum Wells (QW), Wide Quantum
Wells (WQW) ending with three- dimensional bulk
samples. In each case the theoretical description should be
different, since the various relations between the optical
confinement (characterized by the ratio between wave length and
dimension $\frac{\lambda}{L}$), the quantum-mechanical confinement
(the ratio of a size in the growth direction to the effective
exciton Bohr radius) and the coherence length, have to be taken
into account. In the present paper we will discuss
the examples of QDs, QWWs, QWs, WQWs,  and bulk crystals, assuming
in all the cases the cylindrical symmetry. We extend the RDMA to
examine systems with various dimensionality, and in all cases the
analytical expressions for susceptibility will be derived, which
enable one to calculate the absorption spectra. Moreover, in the
bulk system, the role of polaritons, being the superposition of
electromagnetic field and quantum coherence modes, will be
considered and the influence of their relative contribution on
matching the experimental and calculated resonances positions will
be presented.

The paper is organized as follows. In Sec. II we recall the basic equations of the RDMA.  In Sec. III we explicitly derived the formula for susceptibility for quantum dots  while section IV is devoted to detailed analysis of the case of quantum wires.  The formulas derived in Sections II are  applied in Sec. V and VI, which are devoted to presentation of optical properties for Cu$_2$O quantum wells and wide quantum wells.  In Sec. VII we consider the case of bulk crystals, where the optical properties for exciting energies near the fundamental gap  are dominated by exciton-polaritons and show the dispersion relation in such situation. Sec. VIII contains illustrative numerical results while a summary and conclusions of our paper are presented in Sec. IX.

\section{Basic equations}
 We study the optical properties of RE in Cu$_2$O based
low-dimensional systems (LDS).
The lowest considered exciton state is the $2p$ state with the extension of about 4 nm.
%Poszlo do wstepu
We will use the Real Density Matrix
 Approach, applied to systems with reduced dimensionality, showing the phenomenon of Rydberg states.
  In this approach the optical properties are described by
equations for the coherent amplitudes $Y_{12}$ of the
electron-hole pair of coordinates ${\bf r}_1={\bf r}_h$ and ${\bf
r}_2={\bf r}_e$ which for a pair of conduction and valence bands
\begin{equation}\label{Ylinear}
-i(\hbar \partial_t + \Gamma) Y_{12}+H_{eh}Y_{12} ={\bf M}{\bf
E},
\end{equation}
\noindent where $\Gamma$ is a phenomenological damping
coefficient. In the above equation ${\bf
M}(\textbf{r})$ is a smeared-out transition dipole density,
depending on the coherence radius $r_0 =
\left(E_g/2\mu\hbar^2\right)^{-1/2}$; the $E_g$ is the fundamental
gap and $\mu$ is reduced effective mass of the electron-hole pair.
The \textbf{r} is the relative electron-hole distance
\cite{Zielinska.PRB.2016a}. Specific forms of {\bf M}(\textbf{r})
will be defined in subsequent sections.

RDMA, adopted for semiconductors by Stahl, Balslev, and others \cite{Stahl},
is a mesoscopic approach\cite{Magnetoexcitons_2019}
which, in the lowest order, neglects all effects from the
multiband semiconductor structure, so the exciton Hamiltonian
becomes identical to that of the two-band effective mass
Hamiltonian $H_{eh}$, which includes the electron and hole kinetic
energy, the electron-hole interaction potential and the
confinement potentials\cite{Hecktotter_2018}. In
consequence, the two-band Hamiltonian $H_{eh}$ with gap
$E_{g}$ is
\begin{eqnarray} \label{hamilt}
&&H_{eh}=E_{g} +\frac{{\bf
p}_{h}^2}{2m_h} \nonumber\\
&&+\frac{{\bf p}_{e}^2}{2m_{e}}+V_{eh}(1,2) +V_h(1)+V_e(2),
\end{eqnarray}
\noindent where the second and the third terms on the r.h.s. are
the electron and the hole kinetic energy operators with appropriate
effective masses, the  fourth term is the electron-hole
attraction, and the two last terms are the surface confinement
potentials for the electron and hole.  The total polarization of the
medium is related to the coherent amplitude by
\begin{equation}\label{Polar}
{\bf P}({\bf R})=2 \hbox{Re}\int d^3{r}\,{\bf
M}({\bf r}) Y({\bf R},{\bf r})
\end{equation}
where $\textbf{R}$ is the center-of-mass coordinate. This, in turn, is used in Maxwell's field equation
\begin{equation}
c^2\nabla^2 {\bf E(R)} - \epsilon_b \ddot{\bf E} = \frac{1}{\epsilon_0}\ddot{\bf P(R)}.
\end{equation}
The excitonic susceptibility $\chi$ is then given by
\begin{equation}
{\bf P(\omega,\textbf{k})} =\epsilon_0\chi(\omega,\textbf{k}){\bf E}(\omega,\textbf{k})
\end{equation}
where $\omega$ is the frequency of the incident field and the absorption coefficient can be calculated from
\begin{eqnarray}\label{abscoeff1}
{\bf \alpha}=2\frac{\hbar\omega}{\hbar
c}\hbox{Im}\,\sqrt{\epsilon_b+\chi},
\end{eqnarray}
where $\epsilon_b$ is the background dielectric constant. Analysing LDS, we will consider cylindrical symmetry of the system with the $z$ axis parallel to the incident field. Then the constitutive equation (\ref{Ylinear}) for an LDS takes the form
\begin{eqnarray}\label{carrH}
&
&\Biggl\{E_{g}-{\hbar}{\omega}-i{\mit\Gamma}+\frac{p_{ez}^2}{2m_{e}}
+\frac{p_{hz}^2}{2m_{h}} +\frac{{\bf
p}_{e\parallel}^2}{2m_e}+\frac{{\bf p}_{h\parallel}^2}{2m_h}
\nonumber\\
&&+V_{eh}\left[(\hbox{\boldmath$\rho$}_e-\hbox{\boldmath$\rho$}_h),z_e-z_h\right]+
V_e(z_e)+V_h(z_h)\\
&&+
V_e(\hbox{\boldmath$\rho$}_e)+V_h(\hbox{\boldmath$\rho$}_h)\Biggr\}Y(\hbox{\boldmath$\rho$}_e,\hbox{\boldmath$\rho$}_h,
z_e,z_h)\nonumber\\
&&={\bf M}(\hbox{\boldmath$\rho$}_e,\hbox{\boldmath$\rho$}_h,
z_e,z_h){\bf E}({\bf R}),\nonumber
\end{eqnarray}
\noindent where $V_{e,h}(z_{e,h})$ and
$V_{e,h}(\hbox{\boldmath$\rho$}_{e,h})$ are the confining
potentials in the \emph{z}-direction, and in $x-y$ plane,
respectively, while $V_{eh}$ is the electron-hole interaction
potential. The $\rho=\sqrt{x^2+y^2}$ is the radial coordinate. The
excitonic amplitude $Y$, obtained from the Eq.
(\ref{carrH}), is than inserted into Eq.
(\ref{Polar}), giving the polarization and, finally, the
susceptibility, from which all the optical function of the system
considered can be calculated.

\section{Quantum Dots}
Quantum dots (QDs) systems are confined semiconductor
structures which exhibit a fully discrete spectrum due to the size
confinement in all directions. QDs, mostly based on semiconductors
like Si, InAs, GaAs and other II-VI and III-V compounds, have been
largely conducted and interpreted (for review see \cite{Henini}).\\

Among various shapes of QDs (spherical, Gaussian profile, pyramids
etc.) we have chosen the ones characterized by a cylindrical
symmetry, in particular a disk with the symmetry axis $z$, height $L$ and with infinite hard wall
potentials for electrons and holes in the $x-y$ plane at the
radius $R$. The incident electromagnetic wave is linearly polarized in the $x$ direction.
We assume a parabolic confinement
in the $z$ direction and take the lowest electron- and hole
states in this direction.
 To derive the linear
optical properties of quantum dots
 we need the simultaneous solutions of the constitutive interband equation (\ref{carrH})
 and of Maxwell's equations outside and inside the QD, where the excitonic  polarization is given by (\ref{Polar}),
  including the boundary conditions. In that case, constitutive equation (\ref{carrH})
  refers to a 6-dimensional configuration space $({\bf r}_e,{\bf r}_h)$,
 with appropriate boundary conditions (BC) for the motion of electrons and holes,
 whereas the definition of ${\bf P(R)}$ (${\bf R}$ in Maxwell equations refers to the excitonic center-of-mass
  coordinate within
 the QD). The complexity of the problem can be reduced with some simplifying assumptions. When the carriers (electron or hole) differ in their
 effective mass, one of the possible simplification is to
 "$\,$immobilize\,"$\,$ the quasiparticle with a larger mass in the center of the
 dot, and consider the motion of the other quasiparticle.\cite{Chuu} Contrary to the
 case of GaAs QDs where the heavier particle was the hole, in
 Cu$_2$O it is the electron, so the term ${\bf
p}_{e\parallel}^2/{2m_e}$ vanishes in the effective mass Hamiltonian (\ref{carrH}).
%The effect of the confinement of the electron will be accounted for by
%taking appropriate gap energies.
The average position of the electron is in the center of the disk but is free to move in the $z$-
direction. Both assumptions, i.e. of the electron bounded at $z$
axis, and of the infinite potential for the hole, allows us to obtain
analytical expressions for the disk susceptibility. To sum up, in the constitutive equation (\ref{carrH}) we omit the term ${\bf
p}_{e\parallel}^2/{2m_e}$ while the confinement
potentials have the following form
\begin{eqnarray}\label{boundarydisk}
&&V_{e,h}(z_{e,h})=\frac{1}{2}m_{e,h}z_{e,h}^2,\nonumber\\
 &
&V_h(\rho_h)=\left\{ \begin{array}{ll}
0\quad \mbox{for}\quad \rho_h\leq R,\\
\infty\quad \mbox{for}\quad \rho_h>R.
\end{array}\right.
\end{eqnarray}
\noindent We also use the long wave approximation, neglecting the
spatial distribution of the electromagnetic wave within the
quantum disk. The l.h.s. operator in Eq.
(\ref{carrH}) includes two one dimensional harmonic oscillator
Hamiltonians and the 2-dimensional Coulomb Hamiltonian. Therefore
the solution for the amplitude $Y$ is expressed in terms of
eigenfunctions
\begin{eqnarray}\label{expansion}
&&Y_{n_e}=\\
&&=\sum\limits_{N_e,N_h,j,m}
c_{n_eN_eN_h,j,\ell}\psi^{(1D)}_{\alpha_{ez,N_e}}(z_e)\psi^{(1D)}_{\alpha_{hz,N_h}}(z_h)
\psi_{jm}(\rho_h,\phi),\nonumber
\end{eqnarray}
where and $\psi^{(1D)}_{\alpha_{z},N}(z)$ ($N_e$,$N_h$=0,1,...) are the quantum oscillator eigenfunctions for electron and hole, respectively
\begin{eqnarray}\label{eigenf1doscillator}
&&
 \psi^{(1D)}_{\alpha_{z},N_{e,h}}(z)=
 \pi^{-1/4}\sqrt{\frac{\alpha_{z}}{2^N_{e,h} N_{e,h}!}} H_N(\alpha_{z}z)
e^{-\frac{\alpha_{z}^2}{2}z^2}, \nonumber\\
&&
 \alpha_{z} = \sqrt{\frac{m \omega_{z}}{\hbar}},
\end{eqnarray}
 $H_N(x)$ are Hermite polynomials $(N_{e,h}=0,1,\ldots)$ and $m$ is the effective mass.
 In particular, we consider the lowest confinement state $N_e=N_h=0$. The normalized
 eigenfunctions of the 2-dimensional Coulomb Hamiltonian have a different form depending on the sign of the eigenvalue (energy).
  For the negative energy we obtain
 \begin{equation}
\psi_{jm}(\xi,\phi)=C\,\xi^{\vert
m\vert}\,e^{-\xi/2}M\left(m+\frac{1}{2}-\lambda,
2m+1,\xi\right)\frac{e^{im\phi}}{\sqrt{2\pi}},
\end{equation}
where $j$ and $m$ are the principal and magnetic quantum numbers of the excitonic state,
\begin{eqnarray*}
\lambda=\frac{2}{\alpha},\quad \xi=\alpha\rho,\quad \alpha^2=-\frac{2m_h}{\hbar^2}a_h^{*2}E,
\end{eqnarray*}
$M(a,b,z)$ are the Kummer function (confluent hypergeometric
function),\cite{Abramowitz} and $C$ is a normalization factor. The
eigenfunction, due to the no escape BCs, satisfies the equation
\begin{equation}\label{zeros}
\psi_{jm}(\alpha R,\phi)=0,
\end{equation}
giving the eigenenergies $E_{jn},\;j=0,1,\ldots, n=1,2,\ldots$. In the region of positive eigenenergies, one obtains
\begin{equation}
\psi(\xi)=Ce^{-{i}\xi}\xi^{\vert m\vert}M\left(\vert
m\vert+\frac{1}{2}+{i}\frac{1}{\alpha}, 2\vert m\vert+1,2{\rm i}\xi\right).
\end{equation}
We use the transition dipole density in
the form \cite{Magnetoexcitons_2019}
\begin{equation}\label{dipoledensity}
M(\rho,z_e,z_h,\phi)=\frac{M_0}{2\rho_0^3}\rho\,e^{-\rho/\rho_0}\frac{e^{i\phi}}{\sqrt{2\pi}}\delta(z_e-z_h),
\end{equation}
with the integrated strength $M_0$ and the coherence radius
$\rho_0=r_0/a^*$. The coeficient
$M_0$ and the coherence radius $r_0$ are connected through the
longitudinal-transversal energy $\Delta_{LT}$
as\cite{Zielinska.PRB.2016a}
\begin{equation}
(M_0\rho_0)^2=\frac{4}{3}\frac{\hbar^2}{2\mu}\epsilon_0\epsilon_ba^*\frac{\Delta_{LT}}{R^*}\,e^{-4\rho_0}.
\end{equation}
Using the above equations and considering the lowest confinement
energies $N_e=N_h=0$ in the $z$ - direction, we obtain the expansion
coefficients (\ref{expansion}) in the form
\begin{eqnarray}\label{coefficients}
&&c_{n_e00,j,m}=\\
&&=\frac{\langle M(\rho,z_e,z_h,\phi)\vert
\psi^{(1D)}_{\alpha_{ez,0}}(z_e)\psi^{(1D)}_{\alpha_{hz,0}}(z_h)
\psi_{jm}(\rho,\phi)\rangle}{E_{g}+W_{e0z}+W_{h0z}+E_{hjn}-{\hbar}{\omega}-i{\mit\Gamma}}.\nonumber
\end{eqnarray}
Inserting $Y_{n_e}$ (\ref{expansion}) with the above expansion
coefficients into the Eq. (\ref{Polar}) we compute the mean
quantum disk susceptibility.
%We have calculated the susceptibility
%of Quantum Dots, having in mind the experimental results by Lee et
%al.,\cite{Lee} for the two lowest states $j=0,1$ and the lowest
%confinement states in all directions. The energies
%$W_{e0z},W_{h0z}$ were assumed to be equal to the lowest energies
%of a quantum well with thickness $L$. The energies $E_{jn}$ were
%calculated from the equation (\ref{zeros}), where we used certain
%approximations for the zeros positions.
Performing integration in
(\ref{coefficients}) we obtain the following expression for the
quantum disk susceptibility
\begin{eqnarray}\label{chiQD_M}
&&\bar{\chi}_{QD}=72\sum\limits_{j=0,1}\epsilon_b\left(\frac{a^*_h}{L}\right)\frac{\alpha_{ez}\alpha_{hz}}{p}\;\hbox{erf}
\left(\frac{L\sqrt{p}}{2}\right)\nonumber\\
&&\left(1-\frac{8\nu_j\rho_0}{3(\lambda_j+\rho_0)}\right)^2
\left(\frac{m_h}{\mu}\right)\left(\frac{2}{\lambda_j}\right)^4\left[\left(1-\frac{8\nu_j}{3}+\frac{20\nu_j^2}{9}\right)\right]^{-1}\nonumber\\
&&\times
\frac{\Delta_{LT}\exp(-4\rho_0)}{E_g+W_{e0z}+W_{h0z}+E_{hjn}-\hbar\omega-i{\mit\Gamma}} ,
\end{eqnarray}
where
\begin{eqnarray}
&&\alpha_{ez}=\frac{1}{a^*}\sqrt{\frac{m_e}{\mu}}\sqrt{\frac{W_{e0}}{R^*}},\nonumber\\
&&\alpha_{hz}=\frac{1}{a^*}\sqrt{\frac{m_h}{\mu}}\sqrt{\frac{W_{h0}}{R^*}},\nonumber\\
&&p=\frac{1}{2}\left(\alpha_{ez}^2+\alpha_{hz}^2\right),\nonumber\\
&&W_{e0z}=\left(\frac{\pi a_e^*}{L}\right)^2R^*_e,\\
&&W_{h0z}=\left(\frac{\pi a_h^*}{L}\right)^2R^*_h,\nonumber\\
&&\nu_j=\frac{3(2j+3)}{4(2{\mathcal R}-3)},\qquad {\mathcal R}=\frac{R}{a^*_h},\nonumber\\
&&\lambda_j=j+\frac{3}{2}+\frac{3(2j+3)}{4(2{\mathcal R}-3)},\nonumber\\
&&E_{hj}=-\frac{1}{\lambda_j^2}R^*_h,\nonumber
\end{eqnarray}
where $\hbox{erf}\,(z)$ is the error
function.\cite{Abramowitz}Using the above expressions, the QD
absorption can be calculated from the imaginary part of the
susceptibility (\ref{chiQD_M}). It can be easily
seen that the above expressions are valid in the negative
eigenenergies region. The exciton energies $E_{hj}$ include both
the Coulomb energy and the in-plane confinement energy.
\section{Quantum Wires}
The next type of considered nanostructures are the
quantum wires (QWW). In principal they are mostly obtained from intersection of
quantum wells, so their main properties are quite similar to
these of quantum wells. We choose a quantum wire of cylindrical shape with the radius $R$ and the symmetry axis $z$. In the wire geometry, at least in the $x-y$ section, one cannot separate the relative- and the center-of-mass motion, so that the system has a 5-dimensional
configuration space. In such a case it is hard to
solve the RDMA constitutive equations, therefore we use some approximations. As in
the case of QDs, we take advantage of the fact that the effective electron
mass in Cu$_2$O is much greater than the hole mass but the electron is  allowed  to move in the $z$-axis direction.
With this assumption the basic equation in the RDMA approach (\ref{carrH}) takes the form
\begin{eqnarray}\label{constQWW}
&&\biggl[E_{g\ell_en_e}+\frac{p_z^2}{2\mu}+\frac{p_Z^2}{2(m_e+m_h)}+\frac{{\bf
p}_{h}^2}{2m_h}-
\frac{e^2}{4\pi\epsilon_0\epsilon_b\sqrt{r_h^2+z^2}}\nonumber\\
&&+V_h(\textbf{r}_h)-\hbar\omega-{
i}{\mit\Gamma}\biggr]Y(\textbf{r}_h,z,Z)={\bf
M}(\textbf{r}_h,z){\bf E}(Z),
\end{eqnarray}
 with reduced mass in the z direction $\mu^{-1}=m_{ez}^{-1}+m_{hz}^{-1}$. We assume that the field has a component $\mathcal E$ in the $z$ direction and the transition dipole has a component $M$ in the same direction. In what follows we use
scaled variables $\textbf{r}_h=\hbox{\boldmath$\rho$} a^*_h, \quad
z=\zeta a^*_h,\; {\mathcal R}=R/a^*_h$, the confinement
potential $V$ in the form (\ref{boundarydisk}) and the boundary condition is
\begin{equation}\label{BC}
Y(\rho={\mathcal R})=0.\end{equation} We will solve  the QWW
constitutive equation (\ref{constQWW}) in two limiting cases: for
the strong  and the weak confinements.

In the case of strong confinement limit, we assume that the
confinement effects are larger than the Coulomb
attraction and we use a method analogous to that used in
ref.\cite{Magnetoexcitons_2019}, transforming (\ref{constQWW})
into a Lippmann-Schwinger-type equation
\begin{eqnarray}\label{Lippmann1}
&&\biggl\{\kappa^2-\frac{m_h}{\mu}\partial_\zeta^2
\nonumber\\
&&-\left(\frac{\partial^2}{\partial\rho^2}+\frac{1}{\rho}\frac{\partial}{\partial\rho}+\frac{1}{\rho^2}
\frac{\partial^2}{\partial\phi^2}\right)+V(\rho)\biggr\}Y\nonumber\\
&&=\frac{2m_h}{\hbar^2}a^{*2}_h{ M}(\rho,\phi,\zeta){\mathcal
E}+\frac{2}{\sqrt{\rho^2+\zeta^2}}Y,
\end{eqnarray}
where
$$\kappa^2=\frac{E_g-\hbar\omega-i{\mit\Gamma}}{R^*_h}.$$
The equation (\ref{Lippmann1}) can be solved with the help of the
approriate Green's function $G_{n_e}\left(\rho,\rho';\zeta,\zeta';\phi,\phi'\right)$
\begin{equation}
Y=\frac{2m_h}{\hbar^2}a^{*2}_hG{ M}(\rho,\phi,\zeta){\mathcal
E}+G\frac{2}{\sqrt{\rho^2+\zeta^2}}Y,\end{equation} where
\begin{eqnarray}\label{greenqwhskal}
&&G_{n_e}\left(\rho,\rho';\zeta,\zeta';\phi,\phi'\right)\nonumber\\
&&=\frac{1}{2\pi^2{\mathcal
R}^2}\sum\limits_{\ell=0}^\infty\,e^{i\ell(\phi-\phi')}\sum\limits_{n=1}^{\infty}\frac{J_1\left(\frac{x_{1,n}\rho}{\mathcal
R}\right) J_1\left(\frac{x_{1,n}\rho'}{\mathcal
R}\right)}{\left[J_2(x_{1,n})\right]^2} \nonumber\\ &&\times
\int\limits_{-\infty}^{\infty}\,{d} k \frac{e^{{
i}k(\zeta-\zeta')}}{(m_h/\mu)k^2+\kappa_{n_en}^2},\\
&&\kappa_{n}^2=\kappa^2+\left(\frac{x_{1,n}^2}{\mathcal{R}}\right)^2,\nonumber
\end{eqnarray}
$J_1,J_2$ are Bessel functions of 1st and 2nd order, and
$x_{1,n}$ are the zeros of $J_1(x)$. In order to calculate the susceptibility, we choose the following shape for the amplitude
$Y$
\begin{eqnarray}\label{ansatzyh1}
&&Y(\rho,z)=Y_{0}\frac{\sqrt{2}}{\mathcal
R}\left|J_{2}(x_{1,1})\right|^{-1}\nonumber\\
&&\times J_1\left(x_{1,1}\frac{\rho}{\mathcal
R}\right)\exp\left(-\kappa_{1}\sqrt{\rho^2+\zeta^2}\right)
\frac{e^{i\phi}}{\sqrt{2\pi}}.
\end{eqnarray} The coefficient $Y_0$ is obtained from Eq. (\ref{Lippmann1}).
In this case, we use the transition dipole
density in the form
\begin{eqnarray}\label{m12}
&&M(\rho,\phi,\zeta)=\frac{M_{0}}{\rho_{0h}a^{*3}}
\delta(\rho-\rho_{0h})\,\frac{e^{{
i}\phi}}{\sqrt{2\pi}}\delta(\zeta).
\end{eqnarray}
%\textit{with a coherence radius $\rho_{0h}=r_0/a^*_h$ to juz %bylo pod eq 15}
Using the Green's function (\ref{greenqwhskal}), one arrives at the
following expression for susceptibility
\begin{equation}\label{chi_wire_strong}
\chi_{QWW}=\frac{2}{\epsilon_0}\frac{2m_h}{\hbar^2\,a^*_h}\;\frac{MGM}{1-\frac{MGVY}{MY}},
\end{equation}
where $V=2/\sqrt{\rho^2+\zeta^2}$, and
\begin{eqnarray}
&&MY=Y_{0}M_0\left|J_{2}(x_{1,1})\right|^{-1}\nonumber\\
&&\times J_1\left(x_{1,1}\frac{\rho_{0h}}{\mathcal
R}\right)\exp\left(-\kappa_{1}\rho_{0h}\right),\\
&&MGM=\left(\frac{m_h}{\mu}\right)^{3/2}\epsilon_b\frac{\Delta_{LT}}{R^*}\frac{1}{{\mathcal
R}^2}\sum_{n=1}^\infty
\left[\frac{J_1\left(\frac{x_{1,n}\rho_{0h}}{\mathcal
R}\right)}{J_2(x_{1,n})}\right]^2\frac{1}{\sqrt{\kappa_{n}^2}},\nonumber\\
&&MGVY=\frac{4M_0Y_0}{\mathcal
R}\sum\limits_{n=1}^N\biggl[\frac{J_1\left(\frac{x_{1,n}\rho_{0h}}{\mathcal
R}\right)}{\left[J_2(x_{1,n})\right]^2\vert
J_2\left(x_{1,1}\right)\vert}\frac{1}{\kappa_{n}^2}\nonumber\\
&&\times\int\limits_0^1\,du\,J_1\left(x_{1,1}u\right)J_1\left(x_{1,n}u\right)
e^{-{\mathcal R}\kappa_{nn}u}\biggr].\nonumber\end{eqnarray}

In the weak confinement limit we assume that the
exciton-center-of-mass is confined in the $x-y$ plane while the
electron and the hole move upon the action of the screened 3D
Coulomb potential. With the help of these assumptions, the amplitude $Y$ in the
eq. (\ref{constQWW}) takes the form
\begin{equation}
Y(r,R)=\sum\limits_j\sum\limits_N\,c_{jN}\psi_j(r)\Psi_N(R),
\end{equation}
where we adopt the eigenfunctions $\psi_j(r)$ of the 3D
Schr\"{o}dinger equation appropriate for the \emph{p} excitons,
and for a hard wall confinement potential. The eigenfunction $\Psi_N(R)$ has the
form
\begin{eqnarray*}
&&\Psi_{\ell N }({\bf R})=\frac{\sqrt{2}}{R}\frac{1}{\vert
J_{\ell+1}(x_{\ell,N})\vert}\\
&&\times J_\ell\left(x_{\ell,N}
\frac{R_{\perp}}{R}\right)\frac{\exp(i\ell\Phi)}{\sqrt{2\pi}}.\end{eqnarray*}
Then the susceptibility is given by
\begin{eqnarray}\label{susceptiblityQWWHw}
&&\chi_{QWW}=\\
&&=\epsilon_b\sum\limits_{j=2}^J\sum\limits_N\frac{f^{(3D)}_{j}\Delta_{LT}/R^*}{\left(E_{Tj10}
-E-{ i}{\mit\Gamma}_j+W_N\right)/R^*}\langle\Psi_{0 N }({\bf
R})\rangle^2,\nonumber
\end{eqnarray}
where $E_{Tj10}$ are the excitonic resonance energies and
\cite{Zielinska.PRB.2016a}
\begin{eqnarray*}
&&f_{j}^{(3D)}=\frac{32}{3}\left(\frac{j^2-1}{j^5}\right)\exp\left[\rho_0^2\left(\frac{4}{j^2}-1\right)\right],\nonumber\\
&&W_N=\frac{\mu}{M}\frac{x_{0,N}^2}{{\mathcal R}^2},\\
&&\langle\Psi_{\ell N }({\bf R})\rangle=\frac{1}{\pi
R^2}\int_0^{2\pi}d \Phi
 \int_0^{R}R_{\perp} \,dR_{\perp} \Psi_{\ell N }({\bf R}).\\
\end{eqnarray*}
\section{Quantum Well regime}
In the cases of Quantum Wells (QW) and Wide Quantum Wells (WQW) the higher order states can be
obtained when we consider a "2-dimensional" form of the electron-hole potential,
\begin{equation}\label{2dimpotential}
V_{eh}(\rho)=-\frac{e^2}{4\pi\epsilon_0\epsilon_b\rho}.
\end{equation}
We use the coherent amplitudes $Y$ of the form
\begin{eqnarray}\label{expansion2}
&&Y(\rho,z_e,z_h,\phi)=\sum\limits_{j,m}\sum\limits_{N_e,N_h}c_{jmN_eN_h}u_{N_e}(z_e)u_{N_h}(z_h)\nonumber\\
&&\times\frac{e^{im\phi}}{\sqrt{2\pi}}\psi^{(2D)}_j(\rho)
\end{eqnarray}
with confinement functions $u_{N_e}, u_{N_h}$. The $\psi^{(2D)}_{jm}(\rho)$ are the eigenfunctions of the
Schr\"{o}dinger equation with the potential (\ref{2dimpotential}) and have the form
\begin{eqnarray}\label{eigenfunctions}
&&\psi_{jm}=\frac{1}{a^*}\frac{e^{im\phi}}{\sqrt{2\pi}}\nonumber\\
&&\times
e^{-2\lambda\rho/a^*}(4\lambda\rho)^m\,4\lambda^{3/2}\frac{1}{(2m)!}\frac{[(j+2m)!]^{1/2}}{[j!]^{1/2}}\\
&&\times M\left(-j,2\vert m\vert+1,4\lambda\rho\right),\nonumber\\
&&\lambda=\frac{1}{1+2(j+\vert m\vert)},\nonumber
\end{eqnarray}
corresponding to the eigenvalues
\begin{equation}\label{eigenvalues}
\frac{E_{jm}}{R^*}=\varepsilon_{jm}=-\frac{4}{[1+2(j+\vert
m\vert)]^2},
\end{equation}
where $a^*$, $R^*$ are the excitonic Bohr radius and Rydberg energy,
respectively. Note that the energy $E_{jm}$ is usually modified
with a quantum defect $\delta$, which replaces $j$ with
$j-\delta$, \cite{Gallagher} shifting mostly low-$j$ states and better
reflecting the experimental data \cite{Kazimierczuk}. This empirical correction represents a short-range modification of the Coulomb interaction between electron and hole due to the complex band structure of Cu$_2$O. This, in turn, induces deviations of the exciton binding energies. \cite{Schone2016}
Following the computation scheme presented above, we use the dipole density (\ref{dipoledensity}) in the same form as in QDs. In the considered QW regime the typical wavelength of the input electromagnetic wave is much
larger than the QW dimension, so one usually uses the long wave
approximation. Inserting the formulas
(\ref{2dimpotential}-\ref{expansion2}) into the constitutive
equation (\ref{carrH}) and the polarization (\ref{Polar}), we
obtain the effective susceptibility in the form
\begin{eqnarray}\label{chi2D}
&&\chi^{(2D)}(\omega)\nonumber\\
&&=\sum\limits_{j=0}^\infty\sum_{N_e,N_h}
\frac{\epsilon_b\Delta_{LT}a^*
f_{j}^{(2D)}a_{N_e,N_h}}{L(E_g-\hbar\omega+E_{j}+W_{N_e}+W_{N_h}-
i{\mit\Gamma}_{jN_eN_h})},\nonumber\\
&&f_{j1}^{(2D)}\\
&&={48}\frac{(j+1)(j+2)}{\left(j+\frac{3}{2}\right)^5}\frac{1}{(1+2\lambda\rho_0)^8}
\left[F\left(-j,4;3;\frac{4\lambda\rho_0}{1+2\lambda\rho_0}\right)\right]^2,\nonumber\\
&&E_{j}=-\frac{4}{(2j+3)^2}R^*,\nonumber\\
&&a_{N_e,N_h}=\int\limits_{-\infty}^{\infty}u_{N_e}(z)u_{N_h}(z)
dz,\nonumber
\end{eqnarray}
where $W_{N_e},W_{N_h}$ are the eigenvalues of the confinement
eigenfunctions, $F(a;b;c;z)$ is the Gauss hypergeometric
series\cite{Abramowitz}, and the damping coefficients
${\mit\Gamma}_{j N_eN_h}$ should be specified for any set of
quantum numbers.
For these damping constants, we use the model used in \cite{maser2} which includes temperature dependence and the effects of phonon scattering \cite{Stolz,Kitamura}. All results are calculated for cryogenic temperature ($T=10$ K).
Again, assuming the infinite step-like confinement potentials
$V_{e,h}$ for the electrons and the holes,
eigenfunctions of the corresponding Schr\"{o}dinger equation
\begin{equation}
\left(-\frac{\hbar^2}{2m_{e,h}}\frac{d^2}{dz^2}+V_{e,h}\right)u=W_{N_{e,h}}u
\end{equation}
have the form
\begin{eqnarray}\label{confinement}
&&u_{N_{e,h}}(z)=\sqrt{\frac{2}{L}}\sin\left(\frac{N_{e,h}\pi
z}{L}\right)
\end{eqnarray}
 and the eigenvalues are
\begin{eqnarray}\label{confinement_Eig}
&&W_{N_e}=\frac{\mu}{m_e}\left(\frac{\pi a^*}{L}\right)^2N_e^2R^*,\nonumber\\
&&W_{N_h}=\frac{\mu}{m_h}\left(\frac{\pi a^*}{L}\right)^2N_h^2R^*,
\end{eqnarray}
%with the effective electron and hole masses, and the reduced
%exciton mass $\mu$.
$N_{e,h}=1,2,\ldots$. Using the above expressions
 we obtain the final form of the QW effective
susceptibility
\begin{eqnarray}\label{chi2dinf2}
&&\chi^{(2D)}=\left(\frac{a^*}{L}\right)\sum\limits_{j,N}
\frac{\epsilon_b\Delta_{LT}
f_{j}^{(2D)}}{E_g-\hbar\omega+E_{j}+W_{N}-
i{\mit\Gamma}_{jN}},\nonumber\\
&&j=0,1,\ldots,\; N=1,2,\ldots,\\
&&W_N=\left(\frac{\pi a^*}{L}\right)^2N^2R^*.\nonumber
\end{eqnarray}
The imaginary part of (\ref{chi2dinf2}) is used to calculate the QW absorption coefficient.

\section{Wide quantum well regime}\label{sec_wqw}
When the thickness $L$ of the considered QW increases and is larger than the wavelength of
the propagating wave (300 nm), it corresponds to the Wide Quantum
Well  regime. The long wave
approximation cannot be maintained, but we use the slowly varying envelope approximation\cite{RivistaGC,Mott,Scully}
The Maxwell's equation for the relevant electric vector component inside the WQW satisfies the equation
\begin{equation}\label{Maxwell}
\frac{d^2E}{dz^2}+f(z)E=0,
\end{equation}
with
\begin{equation}\label{efzet}
f(z)=\frac{\omega^2}{c^2}\left[\epsilon_b+\chi(z)\right],
\end{equation}
and the susceptibility $\chi(z)$ can be obtained from the
constitutive equation (\ref{carrH}). The Maxwell's equation then reads
\begin{equation}
E(z)=A_1(z)e^{i\beta(z)}+A_2(z)e^{-i\beta(z)},
\end{equation}
where
\begin{equation}\label{betaz}
\beta(z)=\int\left[f(z)\right]^{1/2}\,d z,\quad
A_i=a_i\left[f(z)\right]^{-1/4}.
\end{equation}
\noindent The coefficients $a_i$ are  obtained with the help of the Maxwell
boundary conditions for the electric field. Thus, the field $E$ within the QW allows one to calculate the optical functions in the  analytical form manifesting their dependence on the confinement
shape.

For the infinite confinement potential, using the eigenfunctions
(\ref{confinement_Eig}), the space-dependent
susceptibility has the form
\begin{eqnarray}\label{chi2dinf}
&&\chi(\omega,z)=\\
&&=\frac{2a^*}{L}\sum\limits_{j,N} \frac{\epsilon_b\Delta_{LT}
f_{j}^{(2D)}}{E_g-\hbar\omega+E_{j}+W_{N}-
i{\mit\Gamma}_{jN}}\sin^2\frac{N\pi z}{L}.\nonumber
\end{eqnarray}
By inserting the above susceptibility to the Eq. (\ref{efzet}), one obtains,
\begin{eqnarray}
&&\beta(\omega,z)\approx\frac{1+\frac{1}{2}\chi(z)(\sin^2\frac{N\pi z}{L})^{-1}}{\left[1+\chi(z)\right]^{1/2}}\;k_bz,
\end{eqnarray} where $k_b=\sqrt{\epsilon_b}\frac{\omega}{c}$. 
With the help of  $\beta(\omega,z)$ one is able to calculate the electric field. By introducing the notation
\begin{eqnarray*}\label{wspolczynnikodbicia}
&&r_{\infty}=\frac{1-\sqrt{\epsilon_b}[1+\chi^{(2d)}/\epsilon_b]}{1+\sqrt{\epsilon_b}[1+\chi^{(2d)}/\epsilon_b]},\nonumber\\
&&\Theta= 2L\sqrt{\epsilon_b}[1+\chi^{(2d)}/\epsilon_b]
\end{eqnarray*}
one obtains the effective refraction index $n_{WQW}$
\begin{equation}
n_{WQW}=\frac{1-r_\infty^2e^{i\Theta}-r_\infty(1-e^{i\Theta})}{1-r_\infty^2e^{i\Theta}+r_\infty(1-e^{i\Theta})},
\end{equation}
which allows for the calculation of absorption coefficient
\begin{equation}\label{alpha_WQW}
\alpha_{WQW}=2\frac{\hbar\omega}{\hbar
c}\hbox{Im}\,n_{WQW}.
\end{equation}

\section{The exciton-polariton regime. Generalized ABC conditions}
When the thickness of the slab exceeds largely
the exciton Bohr radius, the system is 3-dimensional
and some new aspects, as compared with the above discussed QWs and
WQWs, should be accounted for. For QWs and WQWs the assumptions of
microscopic boundary conditions for the movement of electrons and
holes, combined with a 2-Dimensional Coulomb potential, were
sufficient. In 3 dimensions case, near the crystal surfaces the quasi-particles move in the repulsing
potential of the surfaces, which can be modelled as a hard wall
potential. At a certain distance of the surfaces the e-h Coulomb
interaction prevails and bound states (excitons) are created. The
interaction of excitons with a propagating wave leads to the
formation of polariton waves. The  combined treatment of the
repulsing potential near the surface and the polariton waves in
the bulk is difficult due the different symmetries of
the surface potentials and the Coulomb potential therefore the effort over decades was devoted to the description of
exciton-polariton waves in the context of their interaction with
crystal surfaces \cite{Schneider2001}. The problem called the Additional Boundary
Conditions (ABC) has appeared with the discovery of polaritons -
joint electromagnetic field-matter quasiparticles, which move in a
medium as a superposition of the field and quantum coherence. The
simplest version of ABC rely on two polariton waves propagating
in the half space geometry. When two polariton waves propagate in
the crystal and one of them is reflected, one has to determine three amplitudes. The classical
electrodynamics yield in this case only two boundary conditions
for the electric and magnetic field. Therefore, an additional boundary condition is needed
to obtain a sufficient number of equations. The first proposal
came from Pekar (Pekar's ABC),\cite{Pekar} which assumed the
polarization to be zero at the crystal surface. His ABC was then
improved by Hopfield and Thomas\cite{Hopfield} who assumed that
the polarization vanishes at a certain surface inside the crystal.

  The ABC problem becomes more complicated when more than 2 polaritons can propagate (i.e., in GaAs
and GaAs based superlattices) or higher excitonic states are involved. Various ABC models, going beyond
the above mentioned, have been proposed for this case.\cite{Andrea},\cite{Birman82}$^-$\cite{Agran2009},\cite{Kalt}. The Rydberg excitons
and polaritons are an exceptional example, in which a  huge  number of polariton waves can appear. It is well-known that Pekar's ABC are applied for an arbitrary surface $(0,L=z)$, but it is assumed that excitons-polaritons appear at the distance of several Bohr radii from the surface. In the case of $j=25$, the critical distance is less then 1$\mu$m, while for states characterized by smaller $j$ these distances are considerably smaller. Therefore the excitons-polaritons might be observed in structures with the quantum confinement effects\cite{Schiumarini2010}, providing fine enough spectral resolution.\cite{Schneider2001}

 Here we propose a certain modification of the
Pekar-Hopfield-Thomas model (PHT) which is applied for the case of Rydberg exciton-polaritons. Since we will assume that the polarization vanishes at the surface, this will correspond to the ,,no escape'' conditions for electrons and holes, defined by equations (\ref{confinement}).  We start with the polariton dispersion relation (taking into account only $P$
excitons)\cite{Nonlinear}
\begin{eqnarray}\label{susceptiblityanisotropic}
&&\frac{k^2}{k_0^2}-\epsilon_b=\chi_{eff}^{(3D)}=\nonumber\\
&&\epsilon_b\sum\limits_{j=2}^N\frac{f^{(3D)}_{j}\Delta_{LT}/R^*}{\left(E_{Tj10}-E-{i}{\mit\Gamma}_j\right)/R^*+(\mu/M_{tot})(ka^*)^2}
\end{eqnarray}
where $M_{tot}$ is the total excitonic mass, and the energies of
exciton resonances $E_{Tj10}$ are known. Note the polaritonic contribution $(\mu/M_{tot})(ka^*)^2$, which shifts the excitonic energies. In order to apply
the PHT model we assume, that for exciting energy near a certain exciton resonance $E_{Tj10}$ the biggest contribution to the optical functions comes from two polariton waves with wave
vectors $k^{(j)}_1, k^{(j)}_2$, which are the two solutions of Eq. (\ref{susceptiblityanisotropic}) nearest to the axis $E=k_b\hbar c/\sqrt{\epsilon_b}$. With them,
we define the partial contributions $\chi_{1,2}^{(j)}=\chi_{eff}^{(3D)}(E_{Tj10},E,k^{(j)}_{1,2})$ to
the susceptibility and, in accordance with the Pekar's model,
 we assume that the contribution to the exciton polarization coming from these two waves
 with amplitudes $E_1^{(j)}, E_2^{(j)}$ vanishes at the crystal surface
\begin{equation}\label{Pekar}
\chi_1^{(j)}E_1^{(j)}+\chi_2^{(j)}E_2^{(j)}=0.
\end{equation}
The above equation, supplemented with the Maxwell's BCs for the
electric field, allows for calculating (in the half-space
geometry) the amplitudes of the polariton waves. This model can be
easily extended to include the polariton waves reflected at the
second crystal surface. The partial susceptibilities define the
indices of refraction of polariton waves, by the relation
\begin{equation}\label{indices}
\left[n_{1,2}^{(j)}\right]^2=\chi_2^{(j)}+\epsilon_b.
\end{equation}
It follows from the above equation that the polariton waves have
different indices of refraction, a property, which can be used in
separating polariton waves propagating through the crystal.

Finally, to calculate the absorption coefficient it is necessary to introduce an additional summation in Eq. (\ref{susceptiblityanisotropic}) over these partial wave vectors $k^{(j)}_1, k^{(j)}_2$, obtaining susceptibility which is then used in Eq. (\ref{abscoeff1}).

\section{Results of specific calculations}

\begin{figure}[h]
\centering
\includegraphics[width=.8\linewidth]{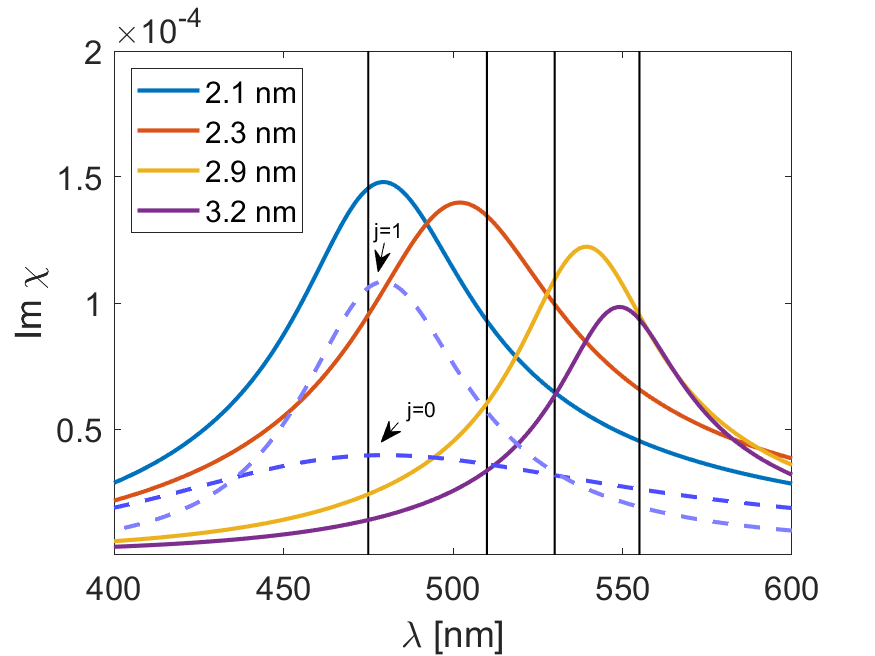}
\caption{Absorption spectra of cylindrical quantum dots calculated
for 4 values of the disk  radius. Black lines mark the
experimental data measured by Lee \emph{et al}.\cite{Lee}}
\label{imchiQD}
\end{figure}

We have calculated the absorption from the imaginary part of
$\bar{\chi}_{QD}$ defined in Eq. (\ref{chiQD_M}), for a Cu$_2$O QD
system and compare our theoretical predictions with the
experimental results by Lee \emph{et al}.\cite{Lee} In the
calculations the two lowest exciton states $j=0,1$ and the lowest
 confinement states in the $z$-direction were accounted for.
The parameters used in calculation are summarized
in Table \ref{parametervalues}. The results are shown on the
Fig.\ref{imchiQD}. Since the experiments in Ref.\cite{Lee} were
performed for spherical dots, we have slightly changed the
dimensions, using an effective radius $R = 4/5 R_{spherical}$ and
the disk height $L = 1.6 R$. One can see that the contributions
from $j=0$ and $j=1$ states (dashed lines) overlap, forming
single, wide absorption maximum. Our calculated theoretical curves
agree very well with the experimental absorption curves from
Ref.\cite{Lee} We observe the increasing blue shift with
decreasing QD radius, and the increasing oscillator strength with
lowering the dimensions. As it was
observed\cite{Borgohain}$^-$\cite{Zhou} for the case of Cu$_2$O
based QD, the excitonic transition energies depend strongly on the
lateral extension.

\begin{figure}[h]
\centering
\includegraphics[width=.85\linewidth]{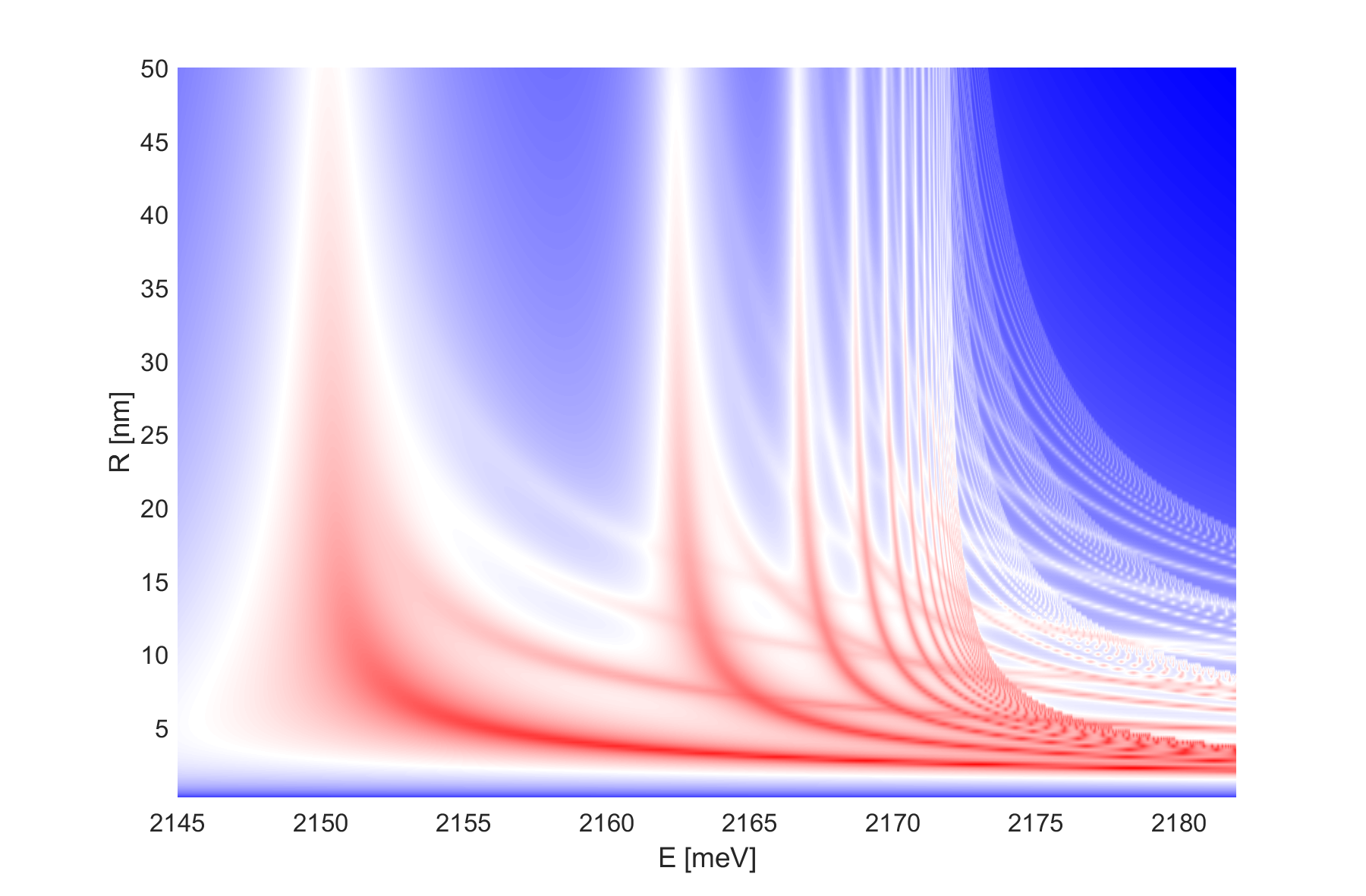}
\caption{The absorption coefficient $\alpha$ of Cu$_2$O nanowire, calculated from (\ref{susceptiblityQWWHw}),
 as a function of the wire radius R.}
\label{A_wire}
\end{figure}

The Fig. \ref{A_wire} shows the absorption coefficient (\ref{abscoeff1}) calculated for the case of a quantum wire, using the susceptibility given by Eq. (\ref{chi_wire_strong}). One can observe multiple excitonic states which diverge towards the higher energy as the wire radius approaches
0 and for small wire radius one may obtain a
strong enhancement of the binding energies.
The confinement states become more visible at low $R$. For sufficiently small radius, these lines mix and overlap, producing a complicated pattern. One can also observe that most of these confinement states are located above the gap energy; lower excitonic states are stronger bounded than the higher ones. These tendencies are in agreement with available experimental and theoretical results for  Rydberg states of excitons in GaAs quantum wires
\cite{okano, banai}.

\begin{figure}[h]
\centering
\includegraphics[width=.85\linewidth]{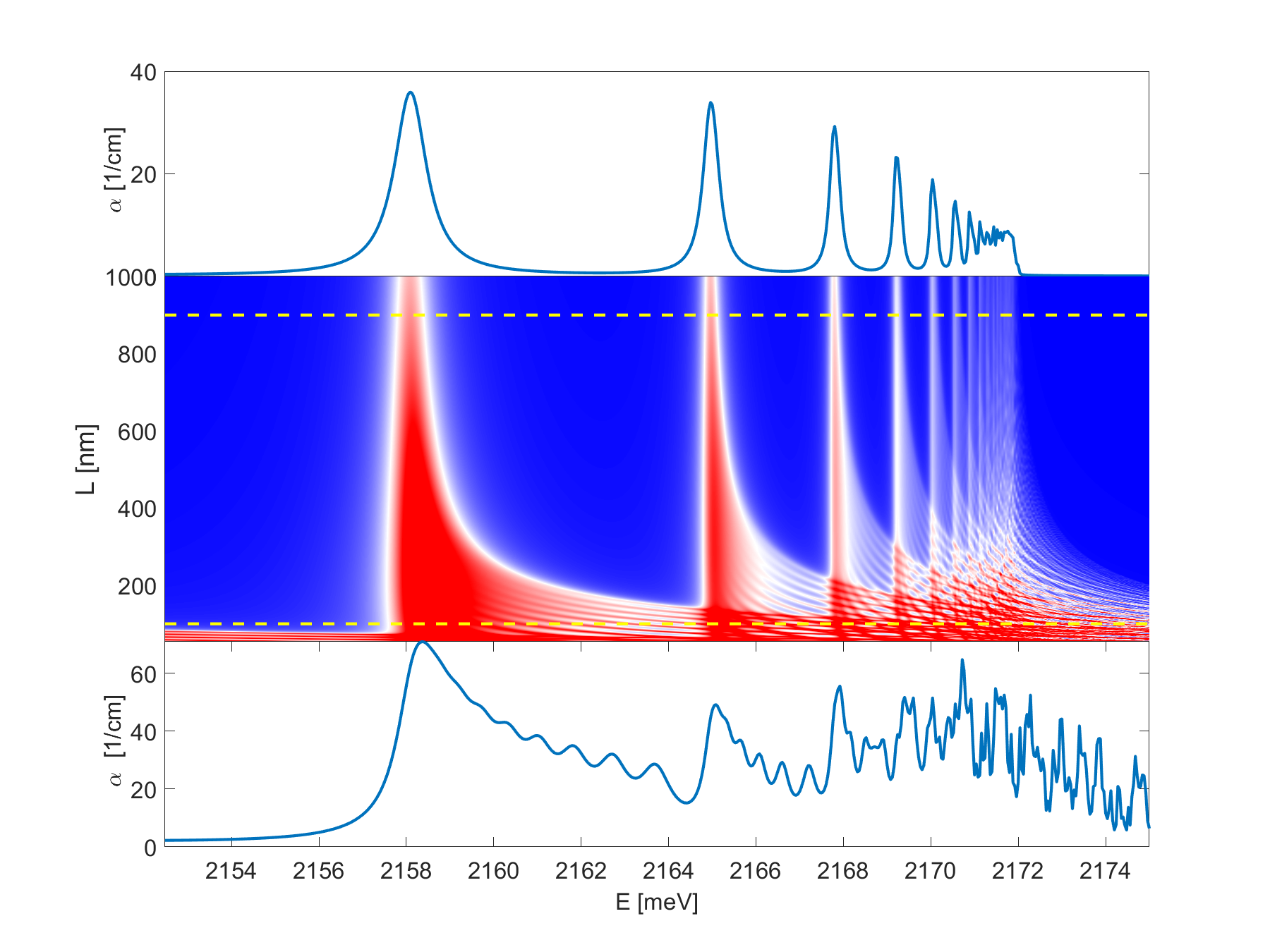}
\caption{The absorption coefficient $\alpha$ of Cu$_2$O crystal, calculated from (\ref{chi2dinf2}),
 as a function of crystal thickness. The top and bottom panels
show cross-sections for thickness $L=900$ nm and $L=100$ nm, respectively.}
\label{imchi2d}
\end{figure}

\begin{figure}[h]
\centering
\includegraphics[width=.9\linewidth]{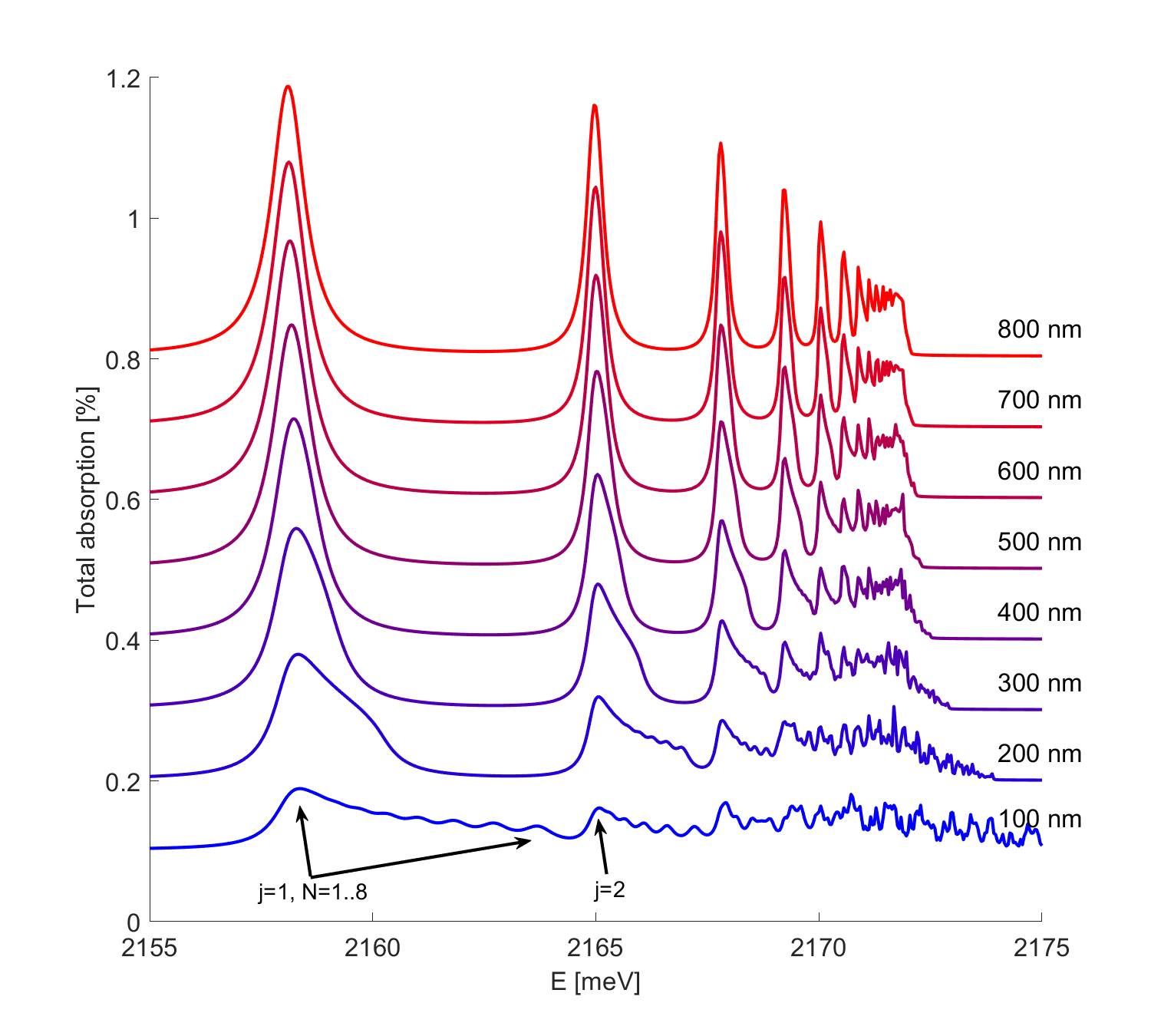}
\caption{The absorption coefficient from Fig. (\ref{imchi2d}), for selected values of thickness.}
\label{imchi2d_2}
\end{figure}

As a next step, we consider a quantum well in the form of a
plane-parallel slabs of Cu$_2$O. In our calculations, the
dimensions in the $z$-direction varied from 20 nm to micrometer
range, which corresponds to structures used in experiments by
Takahata \emph{et al} \cite{Takahata} (lower limit) and  by
Kazimierczuk \emph{et al} \cite{Kazimierczuk} (upper limit).  Such
dimensions cover the regimes of QWs, WQWs, and exciton-polariton
regime. For any regime the calculations were performed by methods
appropriate to the given regime. The limits between these regimes
are not sharply defined. For example, the thickness $L$=200 nm is
large compared to the extension of the lowest exciton state (about
4 nm), but small compared to the extension of states with $j>10$.
Therefore we used the criterion of the relation between the slab
thickness and the wavelength of the wave propagating in the
crystal, which equals to about 200 nm. We consider the slabs with
$L<200$ nm as QWs and use the long-wave approximation, which,
together with the assumption of infinite confinement potentials,
leads to the expression (\ref{chi2dinf2}) for the effective
dielectric susceptibility and, in consequence, to the expression
for the absorption coefficient. The absorption line shape
resulting from Eq. (\ref{abscoeff1}) is shown in the lowest part
of Fig. \ref{imchi2d}. We observe the overlapping of exciton and
confinement states. For small $j$ and
$N$ the exciton effect prevails, whereas for
large values of $N$ the series of exciton
resonances appears below every confinement state. These peaks
exhibit a strong, roughly parabolic shift towards higher energy
with decreasing $L$, which is similar to the case
of quantum wire and typical for these structures in other
semiconductors \cite{Christol}. Eventually, the lines cross and
mix together, creating a complicated spectrum, especially for $E >
E_g$. Interestingly, due to the large number of confinement states
present only in a thin crystal, the absorption coefficient is
decreasing with $L$. However, the total absorption is still
proportional to thickness, as shown in Fig. \ref{imchi2d_2}. One
can also observe that the absorption discontinuity at the band gap
is smeared out and disappears completely at for $L<200$ nm.
The relative amplitude and shape of absorption
peaks and the strong mixing of higher states are consistent with
experimental observations by Khramtsov \emph{et al} for the
GaAs/GaAlAs quantum wells. \cite{Khramtsov} For $L>200$ nm the
long-wave approximation is not valid, and the methods of Sec.
\ref{sec_wqw} are used. The effect of confinement decreases, and
the maxima related to the exciton states with $j=2,3...$ are
visible. This effect is also observed in the central part of Fig.
\ref{imchi2d}. When the considered crystal thickness is
considerably larger that the wavelength inside the crystal, the
reflection and transmission spectra will be strongly influenced by
Fabry-Perot interference. One can observe that the absorption
maxima on the Fig. \ref{imchi2d} exhibit strong mixing for $L<200$
nm, which creates a very complicated transmission pattern.

\begin{figure}[h]
\centering
\includegraphics[width=.9\linewidth]{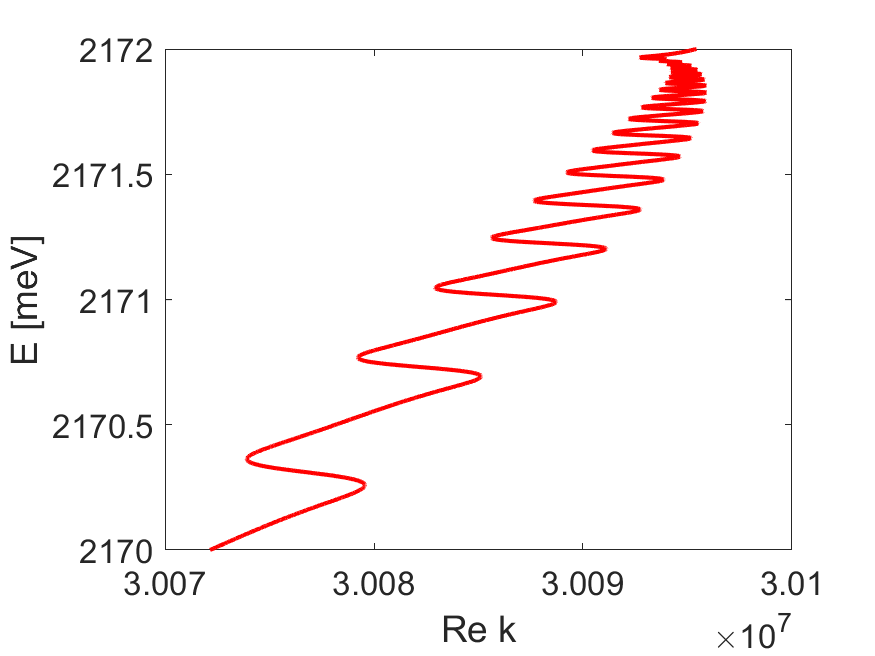}
\caption{Polariton dispersion relation calculated from (\ref{susceptiblityanisotropic}).}
\label{wykres_k}
\end{figure}

\begin{figure}[h]
\centering
\includegraphics[width=.9\linewidth]{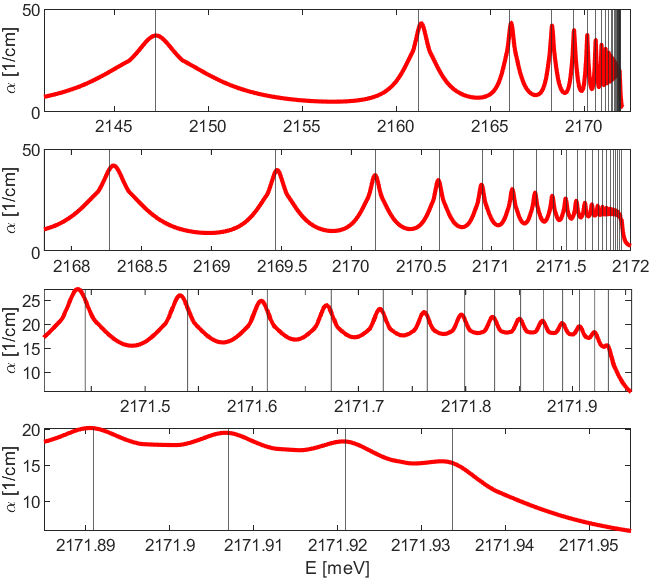}
\caption{Excitonic absorption in Cu$_2$O crystal with polaritons calculated from (\ref{abscoeff1}) and (\ref{susceptiblityanisotropic}).
Black lines are peaks from experimental data by Kazimierczuk\emph{
et} \emph{al}.\cite{Kazimierczuk} }
\label{porownanie_kazimierczuk_abs_comp}
\end{figure}

\begin{figure}[h]
\centering
\includegraphics[width=.85\linewidth]{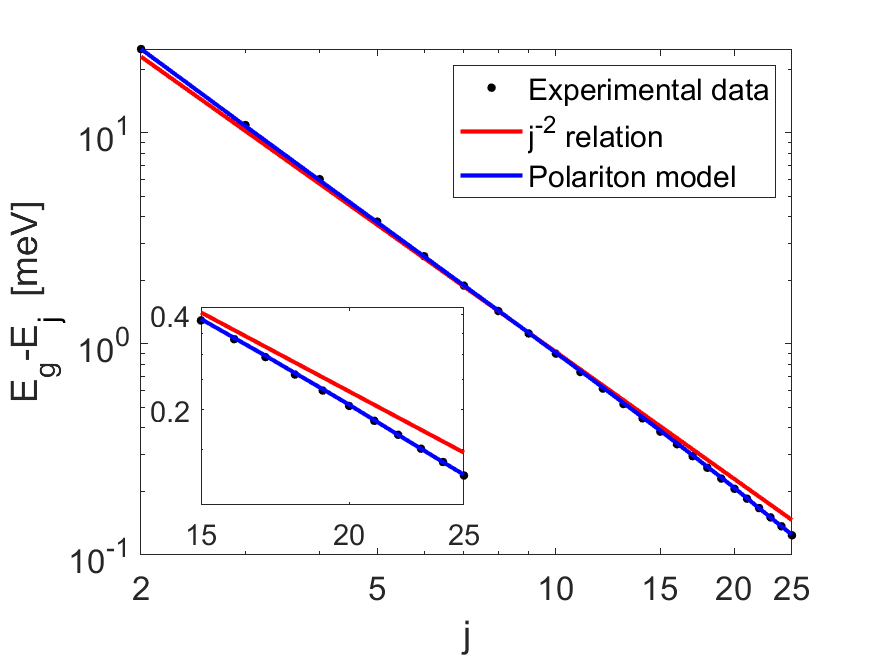}
\caption{Absorption peaks position comparison between theory with (\ref{susceptiblityanisotropic})
or without polaritons to experimental data by Kazimierczuk
\emph{et} \emph{al}.\cite{Kazimierczuk} Line $j^{-2}$ added for
reference. } \label{porownanie_kazimierczuk_n_}
\end{figure}

Fig. \ref{wykres_k} shows the exciton dispersion relation (\ref{susceptiblityanisotropic}),
including polaritonic contribution which is represented by the term $(\mu/M_{tot})(ka^*)^2$ in the denominator.
Overall, the inclusion the polaritons gives a nonlinear shift to
the position of the resonances; the higher states, which are
closer to $E_g$ are more affected. It should be mentioned that
such an effect could explain some discrepances observed in fitting
a simple $j^{-2}$ model to the available experimental data; the
Fig. \ref{porownanie_kazimierczuk_abs_comp} shows the comparison
between absorption maxima positions measured by Kazimierczuk
\emph{et al} and our absorption spectrum calculated from
(\ref{susceptiblityanisotropic}). For our fit, we have used the
Rydberg energy $R=91.5$ meV and quantum defect $\delta=0.083$.
With these values, we have obtained almost perfect fit to all
excitonic peaks for $j=2\ldots 25$. Note that the quantum defect
affects mostly low-energy states but cannot explain the apparent
deviation from $j^{-2}$ relation for high states. This is easily
visible in Fig. \ref{porownanie_kazimierczuk_n_}; even with proper
fitting values, the standard relation, represented by straight
line, cannot fit all the states. On the other hand, the nonlinear
curve provided by polaritonic relation appears to be a much better
fit.

The above  indicated agreement can be understood as an indirect proof for existence of polariton waves.
 Another argument may come from experiment. Just at early stage of the research on Wannier-Mott excitons
  a number of experiments has been performed to manifest the existence of many transverse waves (polaritons)
   with fixed frequency and polarization, distinguished by the index of refraction (see Broser \emph{et al.} \cite{Broser}).
    In particular, for a CdS crystal, Lebedev \emph{et al.} \cite{Lebedev} observed
    the simultaneous transmission of two polariton waves through a wedge shaped crystal and spatially separated them.
     As we have shown above, see Eq. (\ref{indices}),  a similar situation occurs in a Cu$_2$O crystal:
      near any exciton resonance energy there are two polariton waves with wave vectors
      $k^{(j)}_{1,2}$,
      and different indices of refraction,
      propagating through the crystal. So we hope, that a similar experiment, as for the CdS crystal,
       can be performed for a Cu2O crystal, to give an unambiguous proof for the existence of polariton waves.

%\begin{figure}[h]
%\centering
%\includegraphics[width=1\linewidth]{RHT_N_por}
%\caption{Comparison of reflection between Pekar model and Hopfield-Thomas model. Cu$_2$O crystal in air $n=$1.}
%\label{RHT_N_por}
%\end{figure}

%\begin{figure}[h]
%\centering
%\includegraphics[width=1\linewidth]{RHT_N}
%\caption{Comparison of reflection between Pekar model and
%Hopfield-Thomas model. Cu$_2$O crystal in sapphire $n=1.76$.  }
%\label{RHT_N}
%\end{figure}

\begin{table}[ht!]
\caption{\small Band parameter values for Cu$_2$O, masses in free
electron mass $m_0$, $R^*$ calculated from
$(\mu/\epsilon_b^2)\cdot 13600\,\hbox{meV}$,\,
$R^*_{e,h}=(m_{e,h}/\mu)R^*, a^*_{e,h}=(\mu/m_{e,h})a^*$}
\begin{center}
\begin{tabular}{p{.2\linewidth} p{.2\linewidth} p{.2\linewidth} p{.2\linewidth} p{.2\linewidth}}
\hline
Parameter & Value &Unit&Reference\\
\hline $E_g$ & 2172.08& meV& \cite{Kazimierczuk}\\
$R^*$&87.78& meV &\\
$\Delta_{LT}$&$1.25\times 10^{-3}$&{meV}& \cite{Stolz}\\
$m_e$ & 0.99& $m_0$&\cite{Naka}\\
$m_h$ &0.58&  $m_0$&\cite{Naka}\\
$\mu$ & 0.363 &$m_0$&\\
$M_{tot}$&1.56& $m_0$&\\
$a^*$&1.1& nm&\cite{Kazimierczuk}\\
$r_0$&0.22& nm&\cite{Zielinska.PRB}\\
$\epsilon_b$&7.5 &&\cite{Kazimierczuk}\\
${R_e^*}$&239.4&meV&\\
${R^*_h}$&140.25&meV&\\
 ${a^*_e}$&0.4 &nm&\\
 ${a^*_h}$&0.69 &nm&\\
 $\Gamma_j$&3.88/$j^3$ &meV&\cite{Kazimierczuk,maser2}\\
 \hline
\end{tabular} \label{parametervalues}\end{center}
\end{table}
\section{Conclusions}

A theoretical solutions to model absorption spectra of low
dimensional systems with Rydberg excitons in a wide range of
system dimensions are presented. The optical absorption spectra of
Cu$_2$O quantum dots of different sizes, quantum wires and quantum
wells  as well as for a bulk crystal are discussed. For each
systems dimension the calculations were performed by methods
appropriate to the considered regime leading to the analytical
expression  for the susceptibility.
 Results are compared with available experimental data, showing a good agreement and confirming  that quantum confinement effects are evident from a blueshift in the optical absorption.  In particular, the calculated spectra of all low-dimensional systems exhibit a smooth transition to the bulk absorption in the limit of large size of the nanostructure. For bulk crystals presented calculations performed in terms of microscopic boundary conditions for the exciton motion inside the crystal of finite size are in a good agreement with experimental spectra. Thus, we have shown that the existence of polaritons can explain the positions of exciton resonances with a higher accuracy than existing models.

\section*{Acknowledgments} Support from National Science Centre,
Poland (project OPUS, CIREL  2017/25/B/ST3/00817) is greatly
acknowledged.

\end{document}